\documentclass[oneside,twocolumn]{article}

\usepackage[utf8]{inputenc}
\usepackage{authblk}
\usepackage{hyperref}
\usepackage{graphicx}
\providecommand{\keywords}[1]{\textbf{\textit{Index terms---}} #1}
\usepackage{subcaption}
\usepackage{color,colortbl}

\definecolor{Gray}{gray}{0.9}
\definecolor{LightGray}{gray}{0.5}

\begin{document}

  \author[*,1]{Stefano Bennati}

\author[1]{Aleksandra Kovacevic}

  \affil[*]{Corresponding author}
  \affil[1]{HERE Technologies, Privacy by Design, E-mail: NAME.SURNAME@here.com}

  \title{\huge Modelling imperfect knowledge via location semantics for realistic privacy risks estimation in trajectory data.}

  \maketitle

\begin{abstract}
    {
Mobility patterns of vehicles and people provide powerful data sources for location-based services such as fleet optimization and traffic flow analysis.
Location-based service providers must balance the value they extract from trajectory data with protecting the privacy of the individuals behind those trajectories. Reaching this goal requires measuring accurately the values of utility and privacy.
Current measurement approaches assume adversaries with perfect knowledge, thus overestimate the privacy risk. To address this issue we introduce a model of an adversary with imperfect knowledge about the target.
The model is based on \emph{equivalence areas}, spatio-temporal regions with a semantic meaning, e.g. the target's home, whose size and accuracy determine the skill of the adversary.
We then derive the standard privacy metrics of \emph{k-anonymity}, \emph{l-diversity} and \emph{t-closeness} from the definition of equivalence areas.
These metrics can be computed on any dataset, irrespective of whether and what kind of anonymization has been applied to it.
    This work is of high relevance to all service providers acting as processors of trajectory data who want to manage privacy risks and optimize the privacy vs. utility trade-off of their services.
    }
\end{abstract}


\section*{Introduction}\label{sec:introduction}

Trajectories of people and vehicles provide valuable insights into mobility patterns and behaviors that are key resources for location-based services (LBS) such as traffic optimization or fleet management. At the same time, the same insights pose a privacy risk for individuals, as the spatio-temporal nature of location data reveals an additional semantic context about them (e.g. analysis of pick-up and drop-off location and time stamp might reveal home or work addresses). Additionally, location can be linked to large amounts of publicly available data such as phone books or social media; this renders location data especially privacy sensitive. Therefore, privacy-aware LBS providers must carefully balance the trade-off between the utility value (e.g. accuracy and freshness) and privacy value (e.g. re-identification risk) associated to location data. 

  The utility of data depends on the characteristics of the LBS to be provided, the service provider has therefore all the information to estimate utility.
  On the contrary, privacy risks depend as well on the knowledge about one or more target individuals available to an adversary \cite{bonchi2009privacy}; therefore, the service provider can estimate privacy risks only after defining a model of the adversary's knowledge with respect to a specific target.

  The most common model in the literature is an adversary with perfect knowledge. Real-world adversaries cannot achieve perfect knowledge because of (i) errors in associating the knowledge about a target with their data, e.g. due to noise in GPS measurements and (ii) availability of additional information, e.g. accessing protected data like medical records.

  The aim of this paper is therefore to argue for the replacement of the perfect knowledge model, as it overestimates the privacy risk and leads to over-aggressive anonymization, in favor adversaries with imperfect knowledge (or \emph{approximate locators} \cite{dewri2012local}), which are more accurate at re-identifying some locations than others depending on their semantic context.

  We distinguish location data in two broad categories: stay-points, e.g. trip origin and destination, and waypoints, e.g. transit on a highway.
    Staypoints are more likely to be associated to a semantic meaning and purpose than waypoints are, e.g. frequent overnight stays at a specific address or regular visits to a worship center.
  The semantic meaning of location is what leads to re-identification and inference of sensitive information; therefore, the paper focuses on the estimation of privacy risks for stay-points \cite{golle2009anonymity}.
  This real-world example demonstrates why the lack of semantic meaning makes re-identification unlikely: Consider the task to re-identify a driver from a waypoint, for example to deliver a fine for an illegal maneuver.
  First of all, the maneuver must be observed, e.g. by a police patrol. The likelyhood of observation is low to begin with as it is not possible to predict the time and location of an illegal maneuver and therefore to dispatch a patrol to observe the event.
  Assuming the event is observed, re-identification would require to infer additional information about the driver of the vehicle. This information is either gained directly, e.g. by reading the license plate of the vehicle, or through analysis of the trajectory dataset, e.g. commute patterns can reveal who the driver is. Given that we aim to estimate the privacy risk associated with location data, we only consider the second situation.
  The next task is to match the observation to the correct data, which is prone to error as both the data and the observation might contain noise in either the spatial and temporal component.
  Combining the likelihood of successfully completing these steps demonstrates that the likelihood of re-identifying an individual from a waypoint is negligible.
  A counter-example are waypoints that contain semantic information, e.g. driving patterns, that allows to associated a trajectory to a specific individual. A discussion on how to model such situations in the proposed framework is presented in section \ref{sec:discussion}.

  The success chances of an adversary to re-identify a stay-point are characterized by \emph{equivalence areas}, spatio-temporal regions where the adversary is unable to distinguish between trajectories.
  The size and shape of equivalence areas reflect the semantic meaning of locations, which might depend on (i) time of day, (ii) information about the map such as type and location of POIs, urban function (residential vs commercial), population density (iii) the road network that allows access to a location, (iv) relevant locations for a target such as favorite places.
  By properly constructing equivalence areas, different adversaries can be modeled: for example the NYC taxi re-identification study \cite{NYC_reidentification_study} could be modeled by placing equivalence areas at the location and time of paparazzi photos.
  In the event where each data point is contained in its own equivalence area, we obtain the perfect knowledge scenario.

  The advantages of our approach are:
  (i) a natural way of modeling imprecision in the data,
  (ii) flexibility to model realistic adversaries with different knowledge and precision by changing the approach to constructing equivalence areas. This flexibility allows the practitioner to better tailor the choice of anonymization to the effective privacy risk, thereby achieving a more desirable trade-off between privacy and utility.
  (iii) independence on the anonymization algorithm and on whether the data has been anonymized or not,
  (iv) compatibility with previous work which relies on a spatio-temporal subdivision of the map, e.g. \cite{montjoye13_unique_crowd,kondor2018towards},
  (v) from equivalence areas we can derive metrics of \emph{k-anonymity}, \emph{l-diversity} and \emph{t-closeness}.

  We apply this model to the use-case of pick-up drop-off (PUDO) analysis of trajectory data, which considers the relation between the origin and destination of trajectories. We run an analysis on the NYC TCL Trip Record Data \cite{NYC_taxi_data} and compute the privacy risks associated with adversaries with different abilities. The risk is computed using standard privacy metrics of \emph{k-anonymity}, \emph{l-diversity} and \emph{t-closeness} which are commonly used with tabular data \cite{sweeney02_k_anony,machanavajjhala06_l,li07_closen} and location data \cite{gruteser03_anony_usage_locat_based_servic,xue2009location,primault2018long}; we believe an advantage of using standard metrics over introducing new scenario-specific metrics is the ability to compare risks across use-cases.

To the best of our knowledge, this is the first work to provide a consistent framework to define all these standard metrics for trajectory data, both for non-anonymized and anonymized data.
Measuring privacy risks before applying anonymization enables identifying weak, privacy revealing subsets of the data where to apply targeted anonymization. This in turn allows to further optimize the privacy vs. utility trade-off.

\section*{Results}
\subsection*{Definition of Equivalence Areas}
Estimating the privacy risk associated to location data requires understanding its semantic meaning.
We define semantic meanings as polygons in space and time called equivalence areas, such that spatio-temporal points falling into the same equivalence area share the same semantic meaning.
Equivalence areas act as  \emph{quasi-identifiers}, i.e. attributes whose values can be linked to external information to re-identify the target \cite{bonchi2009privacy}, such as background knowledge about the target's habits or behavioral patterns.

At the same time, equivalence areas enable the adversary to infer additional personal information about a target, thus act as \emph{sensitive attributes}.
For example, consider commuting from home to work: one specific adversary, e.g. a neighbor, might know the home address and be interested in deducing the work address from the data; another adversary, e.g. a co-worker, might know the work address and be interested to obtaining the home address from the data.
In both cases, matching the trajectory data with one equivalence area allows to determine the second equivalence area and with it infer personal information about the target.
In the remainder of this paper, we assume that the adversary has background knowledge about the origin of a trip and seeks to re-identify trajectories by inferring their destination.
This assumption primarily simplifies communication about the results, as described in Section Discussion \ref{sec:discussion} the methodology outlined in the following can just as well be used to model the case where the adversary tries to infer origins.

Formally, a \emph{trajectory dataset} is a set of trajectories, $T=\{ \tau_1, ... \tau_n \}$, where a trajectory $\tau = \{ p_1, ... , p_m\}$ is an ordered set of tuples, or \emph{records}; each trajectory is associated to a pseudonym $ID_{\tau}$ that allows to identify all records in that trajectory, e.g. a Vehicle Identifier Number (VIN) or a taxi medallion number.
Each record is defined as $p_i = < t_i, s_i, ID_{\tau}, y_i>$, where $t_i$ is the time component, i.e. a timestamp, $s_i$ is the spatial component, i.e. a set of coordinates, $ID_{\tau}$ is the pseudonym of trajectory $\tau$ to which $p_i$ belongs, and $y_i$ is a set of additional attributes of the record, e.g. speed or taxi fare.

We assume that $y_i$ and $ID_{\tau}$ do not contain any personal information about the individual that created the trajectory, nor any information that allows \emph{direct} re-identification or singling out the individual.
An ID such as the taxi medallion number allows to directly link the trajectory to a specific taxi driver who owns the medallion; therefore, we assume IDs are pseudonymized, e.g. replaced by cryptographic hashes.

\begin{figure}
  \centering
  \begin{subfigure}[t]{0.45\textwidth}
    \centering
  \includegraphics[width=\textwidth]{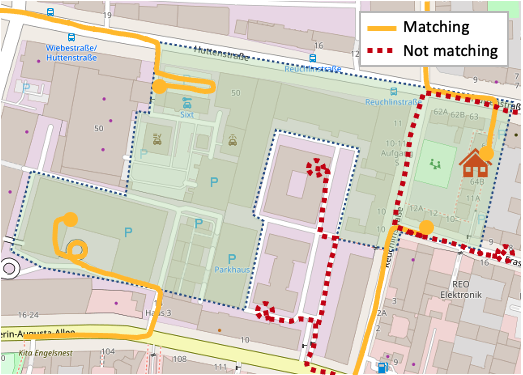}
  \caption{Illustration of trajectories that match (continuous line) and that do not match (dotted line) with the equivalence area (green area). A trajectory matches an area if its starting point (circle) is inside the area.}
  \label{fig:matching}

\end{subfigure}
  \begin{subfigure}[t]{0.45\textwidth}
    \centering
  \includegraphics[width=\textwidth]{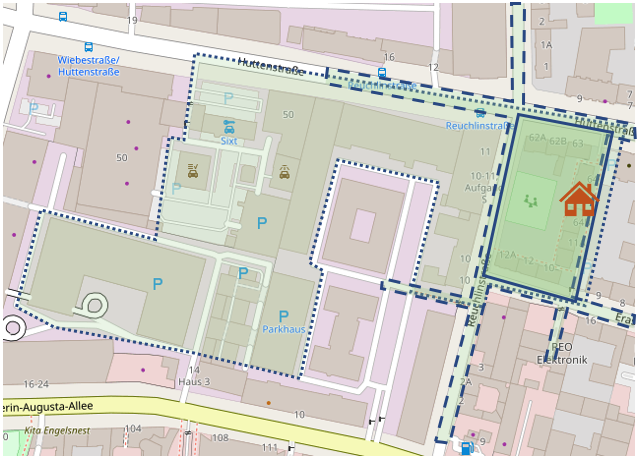}
  \caption{Illustration of equivalence areas obtained by differently skilled adversaries (the temporal aspect of the equivalence areas is not represented): A relatively inaccurate adversary might know the target address but not know their parking habits, thus would consider a large areas covering multiple parking spots where the target might likely park (dotted line); a more accurate adversary could know that the target typically parks on the streets and thus define a smaller area that covers nearby roads (dashed line); a skilled adversary might know that the street where the target lives has reserved parking spots for residents, thus is able to reduce the area down to a block (continuous line).}
  \label{fig:adversaries}
  \end{subfigure}
  \caption{Illustrations of equivalence areas. Map source: OpenStreetMap.}\label{fig:1}
\end{figure}

An \emph{equivalence area} $\mathit{A}$ is one or more (disjoint) polyhedrons in space and time. We denote with $ p_i \in \mathit{A} $ the notion that the coordinates $t_i$ and $s_i$ of a record $p_i$ are contained in the polyhedron (or union of polyhedrons) $\mathit{A}$.
For simplicity, we assume that the origin of a trajectory acts as a quasi-identifier and the destination acts as a sensitive attribute, we thus denote the first and last point of a trajectory as $\tau^{QI}$ and $\tau^{SA}$ respectively.
The set of trajectories that \emph{match with an equivalence area} $\mathit{A}$, i.e. whose quasi-identifier is in the area, is defined as $M_A = \{ \tau \in T : \tau^{QI} \in \mathit{A}\}$ (see Figure \ref{fig:matching}), where $k_A = |M_A|$ is the number of trajectories starting in $\mathit{A}$.

\begin{figure}
  \centering
  \begin{subfigure}[t]{0.45\textwidth}
    \centering

  \includegraphics[width=\textwidth]{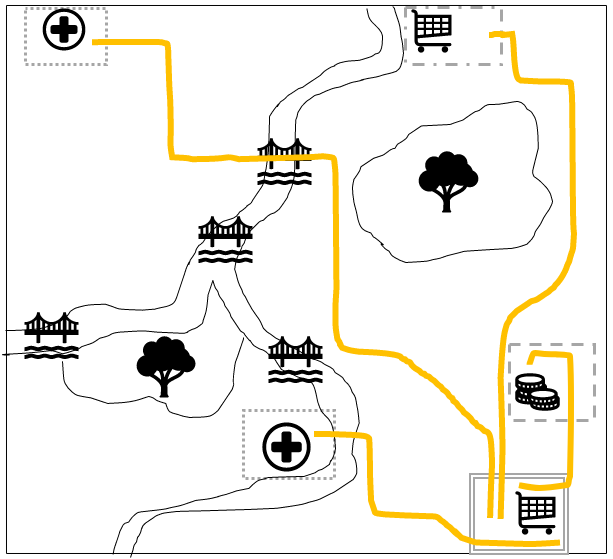}
  \caption{Illustration of a set of inferences for an equivalence area (square with double continuous border). The temporal aspect of the equivalence area is not represented. Four trajectories match with the considered equivalence area, one of them ends in an equivalence area containing a shop (dash-dotted border), one in an equivalence area defined by a bank (dashed border) and two in an equivalence area related to hospitals that are spatially separated but semantically equivalent (dotted border). These three areas form the set of inference.}
  \label{fig:inference}
\end{subfigure}
  \begin{subfigure}[t]{0.45\textwidth}
    \centering
  \includegraphics[width=\textwidth]{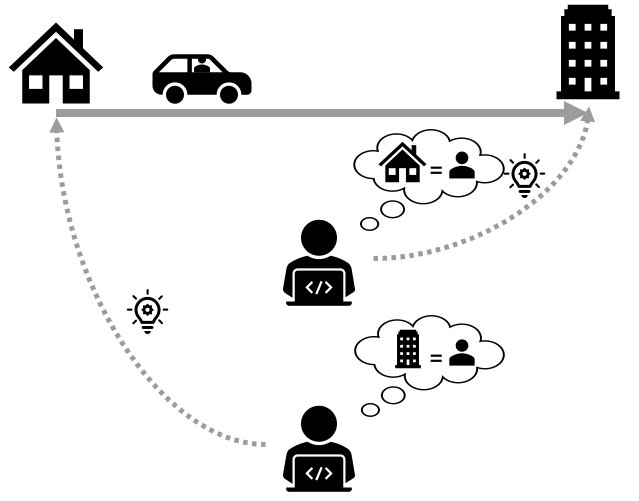}
  \caption{Illustration of how location data can take either the role of sensitive attribute or of quasi-identifier, depending on the knowledge of the adversary: the same trajectory could reveal the target's office (sensitive attribute) to an adversary that knows the target's home (quasi-identifier), or vice versa.}
\label{fig:sensitive_attr}
  \end{subfigure}
  \caption{Illustration of inference on trajectory data and equivalence areas.}\label{fig:2}
  \end{figure}

The set of inferences, i.e. the set of distinct equivalence areas that match with the sensitive attributes of trajectories in $M_A$, is defined as $I_A = \{ \mathit{A}^\prime : \tau^{SA} \in \mathit{A}^\prime, \tau \in M_A \}$ (see Figure \ref{fig:inference}), where $l_A = |I_A|$ is the size of the set of inferences. Each of those equivalence areas is associated with some semantic information, e.g. type of nearby POIs, which can reveal sensitive information about the target.

\subsection*{Model of adversary knowledge}

We consider adversaries with imperfect knowledge, also called \emph{approximate locators} \cite{dewri2012local}, whose ability to re-identify data varies across locations. A typical example is a journalist that is investigating privacy issues in the data, relying on public data such as the address book to obtain a target's home address; the journalist does not have a physical presence that would allow them to tail a target in the real world as they drive.

We model the adversary's ability and background knowledge as the size of equivalence areas, where a more skilled adversary would be able to produce smaller, more accurate equivalence areas.
For example, in a re-identification process, the goal of an adversary is to identify which trajectories in the data are associated to the target with high probability.
  The skill of an adversary can be defined as the precision in identifying these trajectories, which might depend on the availability and the accuracy of background knowledge about a target, e.g. commute habits.
  This background knowledge allows the adversary to define one or more areas which delimit what the adversary believes to be likely associated to the target, e.g. knowing the commute habits of a target, the adversary can limit the analysis to all trajectories that leave the environs of a target's home within a specific time frame.
  A skilled adversary might re-identify all trajectories starting from the street where the target lives and within a 30 minutes time-window of the target's departure, while a less skilled adversary might re-identify all trajectories starting from multiple parking spots within the target's neighborhood (see Figure \ref{fig:adversaries}) and within a 1 hour time-window from the target's departure.

  The maximum accuracy for an equivalence area depends on what resources are available to an adversary, e.g. access to information about POI and population, as well as the intrinsic noise in GPS data. For example, an adversary might pinpoint a target to a specific building but might not be able to distinguish between different semantic meanings associated to the same, e.g. bar vs apartment.
  An equivalence area with an uncertain semantic meaning models probabilistic inference, e.g. the adversary might infer that the trip's purpose is more likely being 'visiting a friend' then 'getting in a bar' except on a Friday night.

Our analysis does not consider the extreme case where the adversary has such a precise knowledge that each data point, including waypoints, falls in its own equivalence area, as this scenario is already considered in previous work.

\subsection*{Experimental evaluation}\label{sec:experiments}

This section presents analysis that we designed to showcase the use of equivalence areas for privacy risk measurement in a real-world use case. The definitions of the privacy metrics of k-anonymity, l-diversity and t-closeness can be found in Section Methods \ref{sec:methods}.

The analysis is based on the NYC TCL Trip Record Data \cite{NYC_taxi_data}, containing information about the location and time of pick-up and drop-off of yellow cabs in the NYC area. This public dataset contains trajectories with only two stay-points: pick up and drop off, so it might not be representative of other applications like parking availability estimation.
We take a subset of data containing the trips that start on Monday 5th of January 2009, although they might end on the next day; around 350,000 trips in total.
Additionally, we clean the dataset by removing trips that last less than a minute or with invalid coordinates, i.e. outside the bounding box defined by the coordinates 74$^{\circ}$ 40'W - 41$^{\circ}$ 30' N and 71$^{\circ}$ 45' W - 40$^{\circ}$ 18' N.

We base our definition of equivalence areas on the NYC Community Districts boundaries \cite{NYC_district_boundaries}, which we enhance by manually drawing additional equivalence areas around the main public transport hubs (airports, train stations, ferry docks). Defining large equivalence areas would likely underestimate the risk caused by a reasonably-skilled adversary, yet community districts have a specific character (business VS residential districts) that allows us to reason about the purpose of trips.
  For the purpose of risk estimation, we recommend to define smaller equivalence areas that reflect the abilities of a realistic adversary.

\begin{figure}
  \centering
  \begin{subfigure}[t]{0.45\textwidth}
    \centering
    \includegraphics[width=\textwidth]{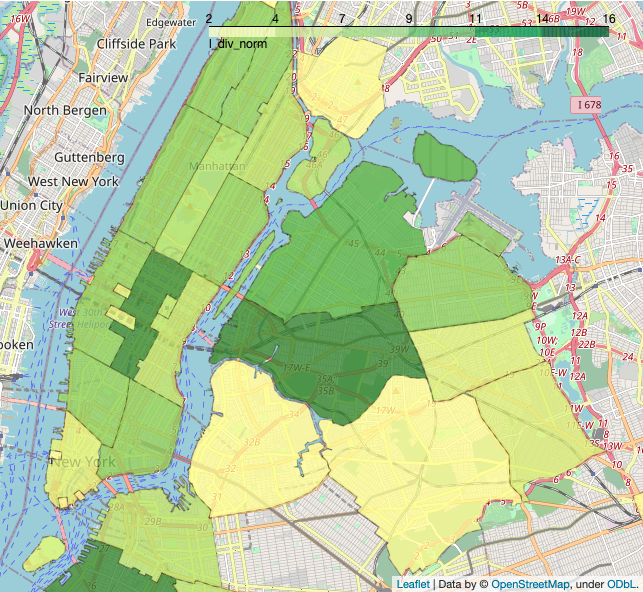}\caption{Trips during the morning rush-hour (7:00-7:30)}\label{fig:heatmap_l_morning}
  \end{subfigure}
  \begin{subfigure}[t]{0.45\textwidth}
    \centering
    \includegraphics[width=\textwidth]{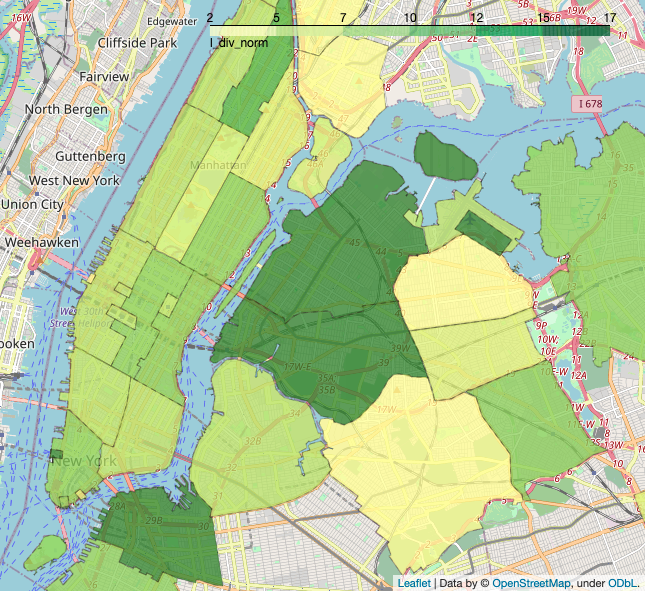}\caption{Trips during the evening rush-hour (17:00-17:30)}\label{fig:heatmap_l_evening}
  \end{subfigure}
  \caption{L-diversity scores for raw data. Lower values mean that fewer destinations are reached by trips starting in that cell. Scores are consistent with commute patterns to and from the main transit hubs and business districts.}
\end{figure}

The first analysis looks at the re-identification risk during the morning rush-hour (between 7 and 7:30, for a total of 4,000 trips) and the evening rush-hour (between 17 and 17:30, for a total of 9,300 trips), which are illustrated respectively in Figures \ref{fig:heatmap_l_morning} and \ref{fig:heatmap_l_evening}. 
The goal of these figures is to highlight differences in the scores caused by different characteristics of the data, for example different commute patterns; for this reason, l-diversity is chosen as score as it captures the flux of trips from one origin to a destination.
When computing this score, we remove trajectories that start and end in the same equivalence areas, which are especially common in Manhattan, as they do not relate to commute patterns.
Lower l-diversity scores mean that fewer destinations are reached by trips starting in that area.

Data shows that a typical commuter pattern is one where taxis transfer workers from the main transit hubs to their offices in the morning and back in the evening. Figure \ref{fig:heatmap_l_morning} shows that the l-diversity score in the main public transit hubs (Penn Station, Grand Central station, Port Authority bus station) is lower than the sorrounding cells, which suggests that most taxi trips that start at these stations end in a few areas, e.g. business districts, while trips starting in the business districts of Manhattan, which might represent business trips, have a more diverse set of destinations. Figure \ref{fig:heatmap_l_evening} shows an opposite pattern in the evenings, where a lower l-diversity in the business districts in Midtown Manhattan indicates that many trips are headed to the major public transit hubs.

\begin{figure}
  \centering
  \begin{subfigure}[t]{0.45\textwidth}
    \centering
    \includegraphics[width=\textwidth]{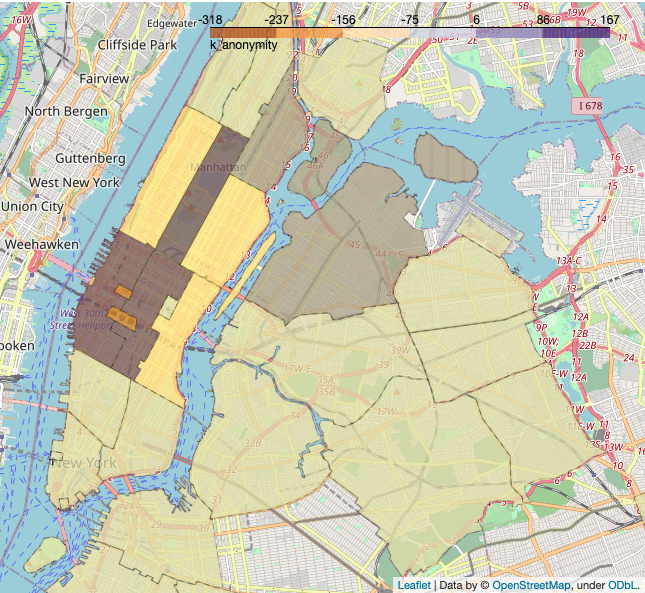}
    \caption{Change of k-anonymity values in the anonymized morning data with respect to the raw data.}
  \end{subfigure}
  \begin{subfigure}[t]{0.45\textwidth}
    \centering
    \includegraphics[width=\textwidth]{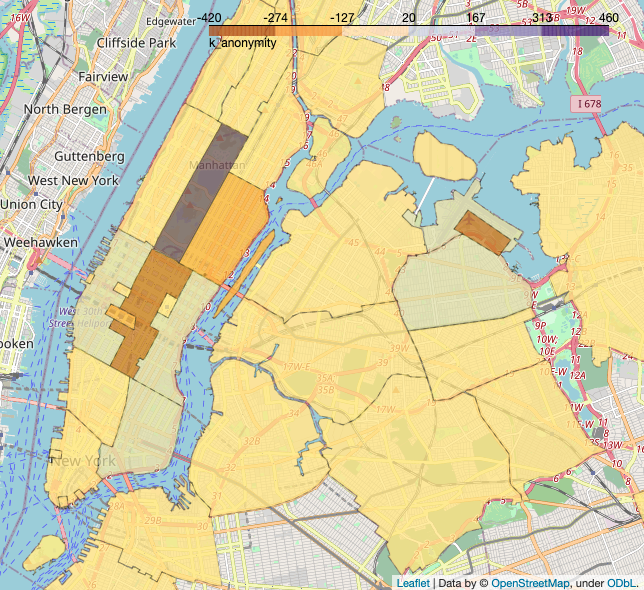}
        \caption{Change of k-anonymity values in the anonymized evening data with respect to the raw data.}
  \end{subfigure}
  \caption{Change in k-anonymity scores after applying anonymization to the dataset.}\label{fig:heatmap_diff}
\end{figure}

The next analysis showcases how the proposed measures can be used to evaluate the effect of an anonymization algorithm: we chose to perturb the location of pick-ups and drop-offs by adding noise sampled from a Gaussian distribution centered in 0 and with a standard deviation of 500 meters; the temporal component has been similarly perturbed with noise sampled from a Gaussian distribution centered in 0 and with a standard deviation of \emph{10 minutes}. The analysis has been repeated 3 times and the scores averaged across repetitions.
Adding random noise to spatio-temporal coordinates has the effect to move these points to nearby locations. If the map is partitioned, e.g. into equivalence areas, perturbed points that are close to the boundaries of these cells have a high likelihood to move to a neighboring cell.
For this reason, we expect anonymization to increase the re-identification risk of dense areas, while decreasing the risk of neighboring areas. This effect is captured in Figure \ref{fig:heatmap_diff}, which shows that k-anonymity values decrease in dense areas such as the main public transit hubs, while increasing in neighboring areas.

\begin{figure}
  \centering
  \begin{subfigure}[t]{0.45\textwidth}
    \centering
  \includegraphics[width=\textwidth]{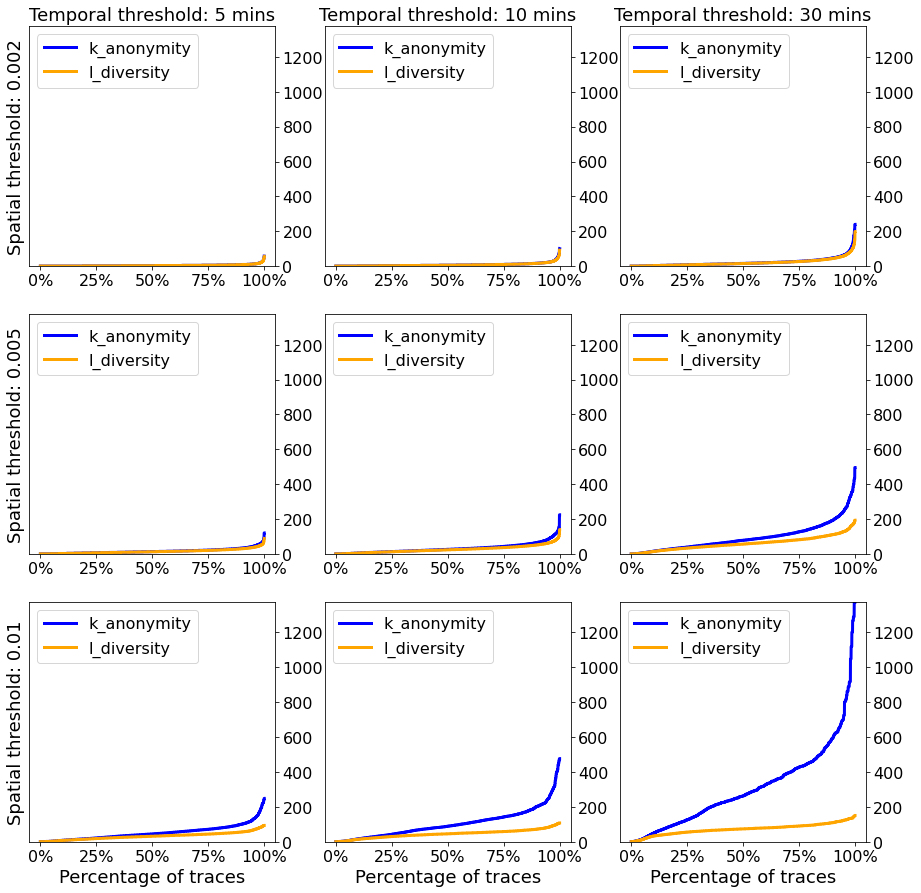}
  \caption{Parameter exploration across equivalence areas of different spatial (rows) and temporal (columns) sizes. Variation of k-anonymity and l-diversity scores. Increasing the size of the areas leads to higher privacy. The leftmost part of the curves represents trajectories with values of k or l below a certain threshold (y coordinate), for example the point (20\%, 3) indicates that 20\% of the traces in the data have less than a score of 3 while 80\% have a score larger than 3.}
  \label{fig:staircase}
  \end{subfigure}
  \begin{subfigure}[t]{0.45\textwidth}
    \centering
  \includegraphics[width=\textwidth]{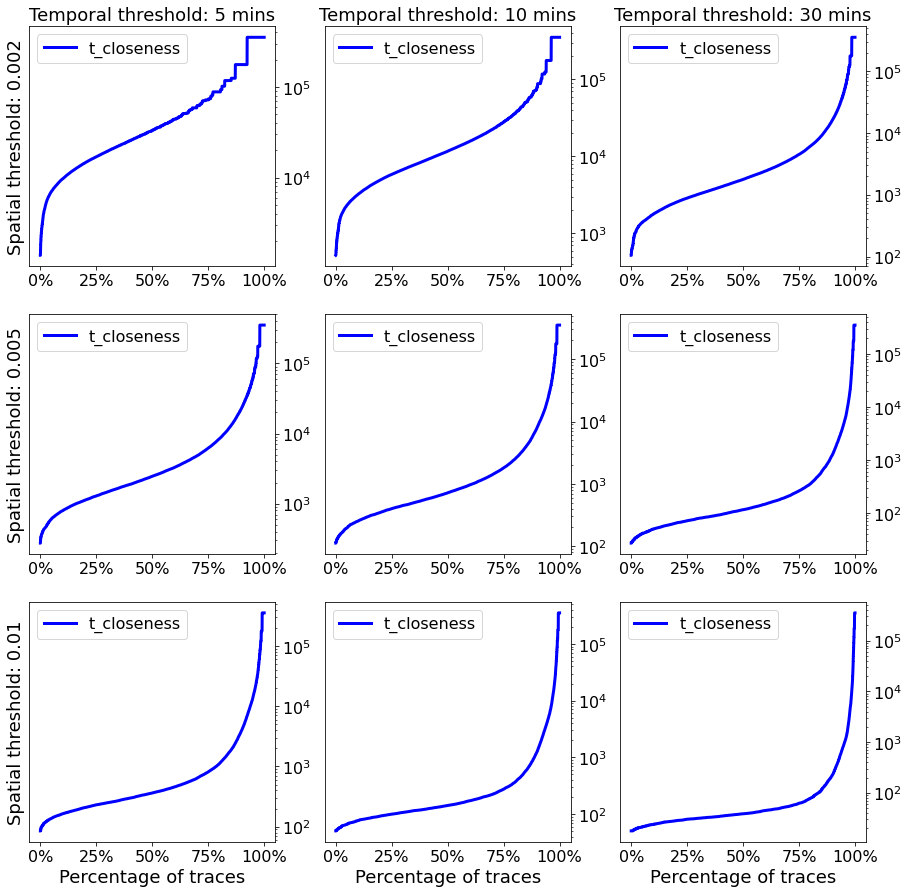}
  \caption{Parameter exploration across equivalence areas of different spatial (rows) and temporal (columns) sizes. Variation of t-closeness scores. Increasing the size of the areas leads to higher privacy, i.e. lower t values.}
  \label{fig:staircase_t}
\end{subfigure}
\end{figure}

The final analysis looks at how the definition of the equivalence areas affects the scores:
For simplicity, we define equivalence areas as cells of a grid with a predefined spatial size (equal for latitude and longitude); the temporal component is similarly partitioned in time-windows of a given size.
The analysis defines equivalence areas for all combinations of a spatial size among 0.002 (corresponding to around 200 meters), 0.005 (corresponding to around 500 meters) and 0.01 (corresponding to around 1 km) and a temporal size among  \emph{5 minutes},  \emph{10 minutes} and  \emph{30 minutes}.
Results are presented on a series of staircase plots, where a score is associated to a percentage of trips in the data. Each row on the plots represents a given spatial size, each column a given temporal size.
Figure \ref{fig:staircase} shows how the scores of k-anonymity and l-diversity change with the size of the equivalence areas: larger areas lead to lower k-anonymity values, as the density increases, while the l-diversity values do not grow as much as an increased grid size reduces the number of possible destinations.

In summary, the analysis shows that the proposed metrics reflect an intuitive understanding of privacy risks: for example, residential areas are associated to a higher risk of re-identification than transportation hubs.
On the contrary, inference risks in transportation hubs are higher as commuting patterns are easier to predict.

\section*{Discussion}\label{sec:discussion}

This work introduces a model of privacy risk estimation for trajectory data which assumes adversaries with imperfect knowledge, e.g. caused by noisy GPS signals or inaccurate background knowledge.
Previous work defines privacy risk as overlap between spatio-temporal regions, which is either computed over spatio-temporal boxes around individual points \cite{montjoye13_unique_crowd,de2015unique} or over spatio-temporal cylinders that surround trajectories \cite{bettini05_protec_privac_again_locat_based_person_ident,abul2008never,nergiz2008towards}.

Both these approaches represent uncertainty as volumes around each data point, while our approach describes uncertainty in terms of \emph{equivalence areas} that correspond to a specific semantic meaning, e.g. associated to a specific concept such as POIs \cite{xue2009location,kondor2018towards,tan2019privacy}, road segments \cite{wang2009privacy}.
One such definitions is location diversity \cite{xue2009location}, an adaptation of l-diversity for location-based queries which measures the diversity in semantic locations where a particular query originates.
A limitation of these and of our model is that they do not consider contextual information that is not related to a specific location, e.g. acceleration profile of the driver, which could lead to re-identification of a trajectory.
  The proposed framework assumes that each trajectory is re-identified via contextual information that is related to one of the stay-points, i.e. the quasi-identifier; in the case the trajectory can be re-identified otherwise, the quasi-identifier should be considered as a sensitive attribute instead.
We believe the proposed framework can be applied to re-identified trajectories, e.g. by considering each stay-point as sensitive attribute for a new virtual trajectory and introducing dummy stay-points to act as quasi-identifier for each virtual trajectory.

Our approach expands on prior work about the measurement of singling-out, linkability and inference in trajectory data. This line of work is based on the framework outlined by \cite{montjoye13_unique_crowd}, where singling-out risk is modeled as the uniqueness of trajectory data. There, data from telco providers is considered, whose location information is represented by pre-defined areas, i.e. related to cellular tower coverage. Uniqueness of trajectories is then computed for different spatial and temporal aggregation of these areas. Differently from our approach, areas are pre-defined and not based on semantic attribute of the locations; additionally, risks of inference and linkability are not considered in the analysis.

  A follow up work takes the same approach to analyze the uniqueness of location data consisting of a series of purchases \cite{de2015unique}.
  Differently from the previous scenario, here location data is associated to a semantic meaning related to the attributes of the transaction, e.g. price, store name. Here, a combination of shop and a time window can be seen as an equivalence area, modeled around an adversary with specific knowledge, e.g. having access to purchase habits of the target but not being capable of standing in the queue behind the target to observe the purchase.
Furthermore, linkability risk is examined by measuring how much the uniqueness of trajectories increases if the adversary knows the approximate price of the transaction, which could be inferred from additional data such as a credit card statement.
The approach presented in our work expands on this by explicitly defining equivalence areas and by proposing a method to derive those from trajectory data which does not already possess a semantic meaning, i.e. GPS coordinates not associated to a specific shop.
The goal of the analysis also differs, as our work examines the risks of singling-out, inference and linkability as a function of the adversary’s background knowledge.

This line of work is further expanded by \cite{kondor2018towards}, where the focus is on measuring matchability, or linkability, between two datasets as the matching points in the second dataset that are within a spatio-temporal radius of the target point. Our proposed method improves on this approach by defining equivalence areas on the semantic properties of the locations, as opposed on a standard value of radius, which enables to estimate singling-out, inference and linkability with a single and uniform processing of the data.

The contributions of this work are
(i) Define a novel threat model for location data privacy in which the adversary has imperfect knowledge about the target.
(ii) Introduce the concept of equivalence areas to model the ability of the adversary at discriminating between locations.
(iii) Introduce a definition quasi-identifiers and sensitive attributes for trajectory data that is based on context and semantics of location data.
(iv) Define singling-out, linkability and inference risks for trajectory data.
(v) Adapt the standard definition of k-anonymity and derivates to measure privacy risks in trajectory data.

Another advantage of our model is the ability to compute the risk on any dataset, irrespective of the anonymization strategy that was applied. This allows the analyst to capture the effect of applying anonymization on the privacy risks. This in turn allows to optimize anonymization by tailoring the anonymization strategy and / or its parameters to specific records or regions based on their intrinsic risks, e.g. apply more aggressive anonymization to the 10\% of traces with the lowest k-anonymity value (those represented in the leftmost part of graphs in Figure \ref{fig:staircase}).

The purpose of the analysis presented in this work is to showcase the use of our model in a real-world setting.
Relying on these results for risk estimation is not recommended due to several limitations of the analysis:

A first limitation is the simplistic definition of equivalence areas, based on static polygons that represent administrative boundaries. This limitation has been mitigated by generating additional polygons around the major public transit hubs, that possess a semantic meaning related to commute patterns. Still, this definition of equivalence areas does not consider map information, e.g. POIs, which is leading to an underestimation of the risk associated to a realistic adversary. An improved definition of equivalence areas could take advantage of existing solutions that define areas around POIs \cite{placekey}.
Generally speaking, the risk estimation obtained from the model refers to an adversary with specific skills, as represented by equivalence areas. A privacy practitioner adopting this model should first evaluate how skilled a typical adversary might be in that context, e.g. evaluating which additional information an adversary might have and the availability of such information, and then define equivalence areas accordingly. The risk estimates obtained with that model would then present an upper bound for the risk associated to any less skilled adversary.

A second limitation is caused by the assumption for simplicity that the adversary has background knowledge about the origin of a target's trajectory and is interested in inferring personal data referring to the trajectory's destination.
This assumption does not limit the validity of the model because the same process can be applied also if these roles are reversed, although it might lead to different values of risk and thus might require to aggregate those.
For example, consider commuting from home to work: one specific adversary, e.g. a neighbor, might know the home address and be interested in deducing the work address from the data; another adversary, e.g. a co-worker, might know the work address and be interested to obtaining the home address from the data. In this example, the home address acts as a quasi-identifier for the first adversary while the work address acts as a sensitive attribute, vice versa for the second adversary (see Figure \ref{fig:sensitive_attr}).

Accurate risk estimation should consider that each stay-point can take either roles. A similar consideration can apply to trajectories with more than two stay-points, in which case the risk could be obtained with the same method by aggregating estimates for each pair of stay-points in the trajectory. Previous work can guide an extension of this model to other use-cases where trajectories typically contain more than two stay-point, e.g where re-identification of purchase transaction data typically requires to match multiple points against a user profile \cite{de2015unique}. Assuming that each point belongs to a different equivalence area, re-identification would require that a trajectory matches on all equivalence areas simultaneously.

Explicitly modeling privacy risks caused by repeating patterns in mobility data, e.g. frequent visits to a worship facility, could also increase the accuracy of estimation of both re-identification and inference risks.
This limitation is driven by the dataset, which lacks information about which trips in subsequent days were associated to the same individual.
In order to model the contribution of patterns to re-identification, we need to explicitly include them in the definition of equivalence areas, e.g. to match all trajectories that start from a specific worship facility at a time that corresponds to the target's habits.
Regarding inference risks, the model can already work with repeating patterns, as repeated visits to a specific location would appear as duplicated values in the sensitive attribute, e.g. multiple destinations in the same equivalence area.

\section*{Conclusions}

To conclude, our work improves on prior efforts by explicitly modeling the (imperfect) adversary knowledge through the definition of \emph{equivalence areas}.
Equivalence areas represent spatio-temporal regions where an adversary, with a given level of knowledge and precision, is unable to re-identify data points related to a target individual, e.g. an adversary with knowledge about the home address of a target might not be able to distinguish the data of the target from that of the neighbors in a given area around the home.
The size and shape of equivalence area reflects the semantic meaning of the location, which depends of features such as nearby POIs, the road network, and so on; as well as the knowledge of the adversary.

Equivalence areas offer flexibility when computing privacy risks, as risks related to adversaries with different capabilities can be computed by adapting how equivalence areas are drawn.

Additionally, we provide a single methodology that works with any type of location data, whether semantic meaning is associated to it in advance or not, whether the data has been previously anonymized or not, and independently of which kind of anonymization algorithm has been used.

Finally, we show how to derive, with a uniform methodology, the standard privacy metrics of \emph{k-anonymity}, \emph{l-diversity} and \emph{t-closeness} from equivalence areas. Using these widely-adopted metrics, it becomes possible to compare privacy risk in trajectory data to risks in other scenarios.
We also show that these metrics are effective to quantify risks of singling-out, linkability and inference using a real-world dataset.

This work can be of interest to data providers and providers of location-based services, e.g. Pick-Up Drop-Off (PUDO) analysis, that want to estimate privacy risk associated to trajectory data, or want to evaluate the effectiveness of an anonymization algorithm.
Data subjects and end users can also find this work relevant to estimating the privacy cost associated with sharing their trajectory data.

\section*{Methods}\label{sec:methods}

\subsection*{Threat model}

The privacy threat comes from an adversary who is looking to extract personal information from a given trajectory dataset. The adversary can either access the data with illegitimate means, e.g. eavesdropping insecure communication (so-called \emph{active} malicious adversary), be authorized to access the data, e.g. the adversary is an \emph{honest-but-curious} service provider \cite{goldreich05_found_crypt_primer}, or the dataset is publicly available.
An example of such adversary is a data scientist or a privacy researcher at a location-based service provider, or a journalist wanting to track a target individual using public mobility data.

A passive adversary can threaten privacy in two ways: (i) Targeting threats: the adversary aims at obtaining additional information about a specific target, i.e. an individual whose data is contained in the dataset and about whom the adversary possesses some background information; (ii) Observing threats: the adversary aims at obtaining additional information about any target of opportunity; the adversary starts by individuating a target with a high privacy risk and only later scouts for publicly available background knowledge about this individual which allows singling out and inference (both privacy aspects are explained in paragraph Privacy Risks \ref{sec:privacyrisks}.)
Our threat model covers both observing and targeting cases of adversary threats.

Therefore, our threat model assumes that: (i) For targeting threats, the adversary knows with certainty that the dataset contains one trajectory created by the target. If this is not the case, the risk of singling-out would be null. For observing threats, we assume that it is possible to find further background knowledge of the (at least one) singled-out individual; (ii) The adversary possesses limited background knowledge about the target. This background knowledge allows the adversary to define a ``semantic region'' that acts as quasi-identifier, e.g. all roads within 5 minutes walking from the target's home on which parking is possible; (iii) The adversary has access to some additional publicly available data sources that allows them to link this background knowledge to regions of the map, e.g. information about POIs, the road network, population density, address directory; (iv) By correctly re-identifying the target's trajectory, the adversary can gain personal information about the target, supplementary to the background knowledge.

Both targeting and observing attacks can be modeled in the same way: (i) The adversary defines an equivalence area based on their background knowledge about the target, which acts as quasi-identifier; (ii) The adversary identifies a set of trajectories that match the target's equivalence area; this result determines the target's singling-out risk, which covers one aspect of the privacy risk; (iii) The adversary performs inference on the values of the sensitive attributes of the matched trajectories; this result determines the target's inference risk, which covers another aspect of the privacy risk; (iv) The adversary links the trajectory data with an external dataset via the semantic meaning associated to equivalence areas, e.g. security camera footage, in order to understand if a trajectory of a specific individual is contained in the dataset.

\subsection*{Definition of Privacy Metrics}
              This section contains a discussion about the choice of privacy metrics for the experimental evaluation.

              The goal of the experimental evaluation is to showcase how a practitioner can use the presented equivalence area concept to measure the privacy risk of location data.
              For this evaluation, we chose the metrics of k-anonymity, l-diversity and t-closeness because their popularity facilitates comparing our results to different use cases.
              A further reason to choose these metrics is the clarity and intuitive understanding about data protection, i.e. the concept of ‘crowd to hide’, that they provide to data subjects, practitioners and regulators.

              Our choice of metrics does not limit the applicability of the framework, the choice of the most appropriate metric for each use case is left to the practitioner that adopts the concept of equivalence areas. As an example, the metric of k-anonymity for location data has been criticized \cite{shokri2010unraveling} as not representative of the actual privacy risk, so a different metric might be more appropriate, e.g. correctness \cite{shokri11_quant_locat_privac}.

              In the remainder of the section, we describe adaptations of the metrics of k-anonymity, l-diversity and t-closeness to scenarios with trajectory data, by relying on the definition of equivalence areas.

In trajectory data a large number of data points forming one trajectory are associated to the same individual, unlike tabular data such as medical records where typically one record is associated to each patient. Adopting the standard definition of k-anonymity, l-diversity and t-closeness would give a wrong estimate of the privacy risk, as the ``crowd to hide'', i.e. data points in one anonymity set, might likely consist of a single individual to which all data points are associated.

\subsubsection*{Privacy Risks}\label{sec:privacyrisks}
Processing personal data involves the following risks \cite{article2014opinion}:
(i) \emph{Singling out} is a risk that a record or a set of records about an individual in dataset can be isolated, without using additional information, e.g. other datasets. This risk is measured by k-anonymity.
(ii) \emph{Inference} is a risk to deduce the value of an attribute from the values of a set of other attributes. For example, access to a healthcare dataset might allow an adversary to deduce the value of diagnosis for a target starting with knowledge about the values of attributes ZIP code and age for that target. Strict k-anonymity measures an upper bound of the inference risk.
(iii) \emph{Linkability} (or \emph{matchability} \cite{kondor2018towards}) is the risk of connecting, across at least two different datasets, data records that corresponds to the same individual. This way the adversary can find more information about the individual than what could be gained from a single dataset, possibly being able to single-out the individual.
While singling out and inference are privacy risks within the given dataset on its own, linkability describes the privacy risks when combining multiple datasets. L-diversity and t-closeness measure each a different aspect of linkability.

\subsubsection*{K-anonymity}
The metric of k-anonymity quantifies the re-identifiability, or singling-out risk, by counting the number of records that are indistinguishable with respect to some background knowledge, i.e. the “crowd to hide”.
If the record identifying the target is in an equivalence area of size $k$, the adversary would not be able to distinguish the correct record from the other $k-1$ false-positives.
We say that each of the $k$ trajectories in the equivalence area described by the background knowledge is \emph{k-anonymous}, has \emph{k-anonymity} $= k$, or that equivalence area is k-anonymous.

\subsubsection*{Strict k-anonymity}
K-anonymity can also measure the risk associated with an adversary finding out whether a trajectory from an observed dataset belongs to a specific individual, based on an additional dataset. For example, an adversary might possess security camera footage showing the individual entering a taxi at a particular location and time. This information can help the adversary to determine if the taxi ride is contained in the observed dataset.
  We cannot know in advance if the adversary has access to data linkable to the equivalence area of the trajectory start, of the trajectory end, or both. Therefore, we take a conservative approach and measure the worst-case scenario of the adversary having knowledge about both origin and destination.
  We measure this linkability risk by computing the \emph{strict k-anonymity}, which is the number of trajectories that are associated to the same equivalence areas at the origin and at the destination: $k_\tau = |\{\gamma \neq \tau: \tau^{QI}, \gamma^{QI}  \in \mathit{A}, \tau^{SA}, \gamma^{SA} \in \mathit{A^\prime} \}|$, where $\tau, \gamma \in T$ are trajectories. This metric measures the ``crowd to hide'' in a more restrictive way than k-anonymity, as it assumes that quasi-identifiers include both the origin and the destination.

\subsubsection*{L-diversity}

K-anonymity covers only one aspect of privacy risk; therefore, a large equivalence area does not necessarily guarantee high privacy, as the sensitive attributes could reveal additional information: the sensitive attributes in an equivalence area of size $k$ might in practice contain only $l<=k$ 'well defined' values. For example, an adversary could infer the work address of a target with certainty if all trajectories leaving from their home address lead to the same office building.
We say that any trajectory that originates in a specific equivalence area $\mathit{A}$ is \emph{l-diverse} if the $k$ trajectories that originate in $\mathit{A}$ end into $l<=k$ distinct equivalence areas. Similarly, we say that equivalence area $\mathit{A}$ is l-diverse.

\subsubsection*{T-closeness}
L-diversity measures how much sensitive information can be disclosed by the sensitive values in an equivalence area; \emph{t-closeness} applies the same concept across equivalence areas, as differences in the distribution of sensitive values in an equivalence area and in the overall population can reveal sensitive information (via a so-called \emph{skewness attack} \cite{li07_closen}).
T-closeness measures the distance between two distributions: $P$, which indicates the frequency sensitive attribute values in the whole dataset, and $Q_A$, which indicates the frequency sensitive attribute values $I_A$ in an equivalence area $\mathit{A}$.
We say that a dataset is \emph{t-close} if the distance $D[Q_A,P]$ is smaller than a threshold $t$, for every equivalence area $\mathit{A}$.
The higher the distance, the more information the adversary can infer from the data; therefore, contrary to k-anonymity and l-diversity, a lower value of \emph{t} means higher privacy.

Figure \ref{fig:staircase_t} shows how the score of t-closeness changes with the size of the equivalence areas: larger areas lead to a steeper curve with more trips having a lower t-closeness value. This is consistent with the findings about k-anonymity and l-diversity (see Figure \ref{fig:staircase}) as it indicates a lower diversity between the contents of equivalence areas and therefore higher privacy.

\bibliographystyle{abbrv}
\bibliography{main.bib}

\end{document}